\documentclass[aps,prb,twocolumn,showpacs]{revtex4}

\usepackage{graphicx}
\usepackage{amsmath}
\usepackage{amssymb}

\begin{document}

\title{Charge dynamics in doped Mott insulators on a honeycomb lattice}

\author{Xixiao Ma}

\affiliation{Department of Physics, Beijing Normal University, Beijing 100875, China}

\author{Yu Lan}

\affiliation{College of Physics and Electronic Engineering, Hengyang Normal University, Hengyang 421002, China}

\author{Ling Qin}

\affiliation{College of Physics and Engineering, Chengdu Normal University, Chengdu 611130, China}

\author{Shiping Feng}


\affiliation{Department of Physics, Beijing Normal University, Beijing 100875, China~~~~}

\begin{abstract}
Within the framework of the fermion-spin theory, the charge transport in the doped Mott insulators on a honeycomb lattice is studied by taking into account the pseudogap effect. It is shown that the conductivity spectrum in the low-doped regime is separated by the pseudogap into a low-energy non-Drude peak followed by a broad midinfrared band. However, the decrease of the pseudogap with the increase of doping leads to a shift of the position of the midinfrared band towards to the low-energy non-Drude peak, and then the low-energy Drude behavior recovers in the high-doped regime. The combined results of both the doped honeycomb-lattice and square-lattice Mott insulators indicate that the two-component conductivity induced by the pseudogap is a universal feature in the doped Mott insulators.
\end{abstract}
\pacs{74.72.Kf, 74.25.Gz, 74.25.F-}


\maketitle

\section{Introduction}

In a Mott insulator, the spin configuration arrangement attempts to minimize the energy by aligning the spins on the lattice sites. However, in some cases, the spin configuration arrangement makes it impossible to minimize the energy on all spins on the lattice sties at the same time - a situation known as the geometrical spin frustration. The parent compounds of cuprate superconductors are a Mott insulator on a square lattice \cite{Bednorz86}, where there is no geometrical spin frustration. Superconductivity is then realized when charge carriers are doped into this parent Mott insulator, where the phase diagram is characterized by two salient phenomena, the high temperature superconductivity and the occurrence of the pseudogap \cite{Hufner08}. In particular, the pseudogap is particularly obvious in the underdoped regime, where the doping concentration is too low for the optimal superconductivity. A number of experimental probes \cite{Timusk99,Batlogg94} show that below a characteristic temperature $T^{*}$, which can be well above the superconducting (SC) transition temperature $T_{\rm c}$ in the underdoped regime, the physical response of cuprate superconductors can be interpreted in terms of the formation of a pseudogap by which it means a suppression of the spectral weight of the low-energy excitation spectrum. In this case, an important issue is whether the pseudogap phenomenon occurred in the doped square-lattice Mott insulators is universal or not. By universality we refer to the pseudogap phenomenon that does not depend on the details of the geometrical spin frustration. The doped Mott insulators on a honeycomb lattice in many ways provide an ideal arena in which to explore this question. This follows a fact that the honeycomb lattice is a loosely coupled lattice with only three nearest-neighbor (NN) sites, where the strongly geometrical spin frustration and quantum fluctuation exist \cite{Shamoto98,Weht99,Kataev05,Moller08,Okumura10,Yan12,McNally15}, and therefore allowing a test of the effect of the geometrical spin frustration on the pseudogap state. Although there are only a few examples of materials where the spins are located on a honeycomb lattice, many fascinating phenomena have been reported in experiments \cite{Shamoto98,Weht99,Kataev05,Moller08,Okumura10,Yan12,McNally15}, such as the quantum spin-liquid state and superconductivity. On the other hand, the Mott insulators on a honeycomb lattice are also of interests in their own right, with many unanswered questions \cite{Sorella12,Murray14,Schaffer14,Zhu14,Jiang14}.

In the doped square-lattice Mott insulator, the physical quantity which most evidently displays the signature for the pseudogap is the charge transport, which is manifested by the optical conductivity and resistivity, where the characteristic features are that the conductivity is divided into two-components by the pseudogap with a non-Drude-like narrow band centered around energy $\omega\sim 0$ followed by a broadband centered in the midinfrared region, while the resistivity grows nearly linearly with temperature \cite{Timusk99,Batlogg94}. In particular, this two-component conductivity extends to the pseudogap boundary in the phase diagram at $T^{*}$. However, this anomalous structure of the conductivity spectrum in the pseudogap state can not be described by the standard Landau Fermi-liquid theory. In our earlier studies \cite{Qin14}, the charge transport in the doped square-lattice Mott insulators has been discussed by considering the pseudogap effect, and then all main features of the optical measurements on the doped square-lattice Mott insulators are qualitatively reproduced \cite{Timusk99}. In this paper, we try to study the charge transport in the doped honeycomb-lattice Mott insulators along with this line. We show that the qualitative behavior of the conductivity in the doped honeycomb-lattice Mott insulators is similar with that obtained in the doped square-lattice Mott insulators. In particular, we show that the transfer of the part of the low-energy spectral weight in the conductivity in the low-doped regime to the higher energy region to form the midinfrared band can be also attributed to the effect of the pseudogap on the infrared response in the doped honeycomb-lattice Mott insulators.

The paper is organized as follows. The basic formalism is presented in Sec. \ref{Formalism}, where we generalize the theory of the charge transport from the previous case in the doped square-lattice Mott insulators to the present case in the doped honeycomb-lattice Mott insulators, and then evaluate explicitly the conductivity. Within this theoretical framework, the quantitative characteristics of the conductivity in the doped honeycomb-lattice Mott insulators are discussed in Sec. \ref{transport}. Finally, we give a summary in Sec. \ref{conclusions}.

\section{Formalism}\label{Formalism}

\begin{figure}[h!]
\centering
\includegraphics[scale=0.5]{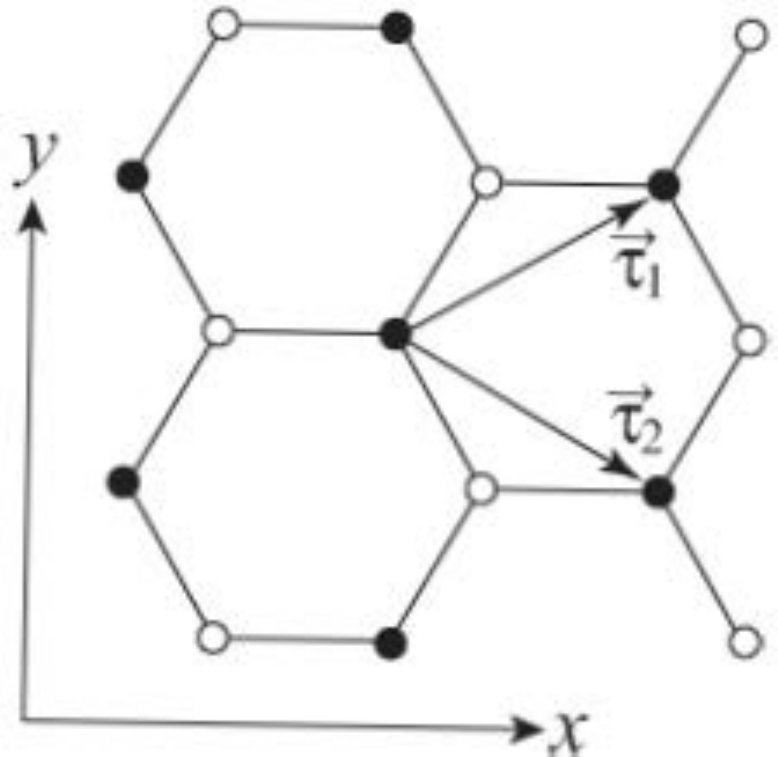}
\caption{The honeycomb lattice with the primitive lattice vectors ${\bf \tau}_{1}$ and ${\bf \tau}_{2}$. Each unit cell contains two sites (indicates by open and solid circles) belonging to the sublattices A and B, respectively. \label{honeycomb-lattice}}
\end{figure}

The $t$-$J$ model, or equivalently the large $U$ Hubbard model, are prototypes to study the strong correlation effects in solids, especially in connection with the doped Mott insulators \cite{Anderson87,Phillips10}. On the other hand, the honeycomb lattice as shown in Fig. \ref{honeycomb-lattice} is formed by the primitive lattice vectors \cite{Sorella12},
\begin{eqnarray}\label{lattice-vectors}
{\bf \tau}_{1}=[3/2,\sqrt{3}/2], ~~~ {\bf \tau}_{2}=[3/2,-\sqrt{3}/2],
\end{eqnarray}
where the lattice constant has been set as $a=1$. Since each unit cell contains two sites, belonging to different sublattices $A$ and $B$, the corresponding reciprocal lattice vectors in momentum-space are given by,
\begin{eqnarray}\label{reciprocal-lattice-vectors}
{\bf k}_{1}=[1/3,1/\sqrt{3}]2\pi, ~~~ {\bf k}_{2}=[1/3,-1/\sqrt{3}]2\pi.
\end{eqnarray}
In this case, the $t$-$J$ model on a honeycomb lattice can be expressed as,
\begin{eqnarray}\label{tjham}
H&=&-t\sum_{l\hat{\eta}\sigma}(C^{\dagger}_{{\rm A}l\sigma}C_{{\rm B}l+\hat{\eta}\sigma}+C^{\dagger}_{{\rm B}l\sigma}C_{{\rm A}l+\hat{\eta}\sigma})+\mu\sum_{\upsilon l\sigma} C^{\dagger}_{\upsilon l\sigma}C_{\upsilon l\sigma} \nonumber\\
&+&J\sum_{l\hat{\eta}}({\bf S}_{{\rm A}l}\cdot {\bf S}_{{\rm B}l+\hat{\eta}}+{\bf S}_{{\rm B}l}\cdot {\bf S}_{{\rm A}l+\hat{\eta}}), ~~~~~~~
\end{eqnarray}
supplemented by a local constraint $\sum_{\sigma}C^{\dagger}_{\upsilon l\sigma} C_{\upsilon l\sigma}\leq 1$ to remove double electron occupancy, where the summation is over all sites $l$, and for each $l$, over its NN sites $\hat{\eta}$, $\upsilon = A, B$ is the sublattice index, $C^{\dagger}_{\upsilon l\sigma}$ ($C_{\upsilon l\sigma}$) is the electron creation (annihilation) operator, ${\bf S}_{\upsilon l}=(S^{\rm x}_{\upsilon l},S^{\rm y}_{\upsilon l}, S^{\rm z}_{\upsilon l})$ are spin operators, $t$ is the NN hopping integral, $J$ is the NN spin-spin antiferromagnetic (AF) exchange coupling constant, and $\mu$ is the chemical potential. It has been shown that the local constraint of no double electron occupancy in the $t$-$J$ model (\ref{tjham}) can be treated properly within the fermion-spin theory \cite{Feng9404,Feng15}, where the constrained electron operators  $C_{\upsilon l\uparrow}$ and $C_{\upsilon l\downarrow}$ are decoupled as,
\begin{eqnarray}\label{fermion-spin-theory}
C_{\upsilon l\uparrow}=h^{\dagger}_{\upsilon l\uparrow}S^{-}_{\upsilon l},~~~~C_{\upsilon l\downarrow}=h^{\dagger}_{\upsilon l\downarrow}S^{+}_{\upsilon l},
\end{eqnarray}
respectively, with the fermion operator $h_{\upsilon l\sigma}=e^{-i\Phi_{l\sigma}}h_{\upsilon l}$ that keeps track of the charge degree of freedom together with some effects of spin configuration rearrangements due to the presence of the doped charge carrier itself, while the spin operator $S_{\upsilon l}$ represents the spin degree of freedom, and then the local constraint of no double electron occupancy is always satisfied in actual calculations. In this fermion-spin representation (\ref{fermion-spin-theory}), the original $t$-$J$ model (\ref{tjham}) can be rewritten as,
\begin{eqnarray}\label{cssham}
H&=&t\sum_{l\hat{\eta}}(h^{\dagger}_{{\rm B}l+\hat{\eta}\uparrow}h_{{\rm A}l\uparrow}S^{+}_{{\rm A}l}S^{-}_{{\rm B}l+\hat{\eta}}+h^{\dagger}_{{\rm B}l+\hat{\eta}\downarrow}
h_{{\rm A} l\downarrow}S^{-}_{{\rm A}l} S^{+}_{{\rm B}l+\hat{\eta}}\nonumber\\
&+&h^{\dagger}_{{\rm A}l+\hat{\eta}\uparrow}h_{{\rm B}l\uparrow}S^{+}_{{\rm B}l}S^{-}_{{\rm A}l+\hat{\eta}} +h^{\dagger}_{{\rm A}l+\hat{\eta}\downarrow}h_{{\rm B}l\downarrow} S^{-}_{{\rm B}l} S^{+}_{{\rm A}l+\hat{\eta}})\nonumber\\
&-&\mu\sum_{\upsilon l\sigma}h^{\dagger}_{\upsilon l\sigma}h_{\upsilon l\sigma}+J_{{\rm eff}}\sum_{l\hat{\eta}}({\bf S}_{{\rm A}l}\cdot{\bf S}_{{\rm B}l+\hat{\eta}}
+{\bf S}_{{\rm B}l}\cdot{\bf S}_{{\rm A}l+\hat{\eta}}),~~~~~
\end{eqnarray}
with $J_{{\rm eff}}=(1-\delta)^{2}J$, and
$\delta=\langle h^{\dagger}_{\upsilon l\sigma}h_{\upsilon l\sigma}\rangle=\langle h^{\dagger}_{\upsilon l}h_{\upsilon l}\rangle$ is the doping concentration.

Within the framework of the fermion-spin theory (\ref{fermion-spin-theory}), the charge transport in the doped square-lattice Mott insulators has been discussed \cite{Qin14}, and the result shows that the striking behavior of the low-energy non-Drude peak and unusual midinfrared band is closely related to the emergence of the pseudogap. In the following discussions, we generalize these analytical calculations from the case in the doped square-lattice Mott insulators \cite{Qin14} to the case in the doped honeycomb-lattice Mott insulators, and then evaluate explicitly the conductivity. However, as we have mentioned above, there are two sublattices $A$ and $B$ in the unit cell, and in this case, the charge-carrier and spin Green's functions are two-dimensional matrices. Following our previous discussions for the case in the doped square-lattice Mott insulators \cite{Qin14}, the corresponding
intra-sublattice and inter-sublattice parts of the full charge-carrier Green's functions can be evaluated explicitly as (see the Appendix \ref{app}),
\begin{subequations}\label{full-charge-Green-function-2}
\begin{eqnarray}
g_{\rm intra}({\bf k},\omega)&=&{1\over 2}\sum_{\nu=1,2}{1\over \omega-\xi^{(\nu)}_{\bf k}-\Sigma^{({\rm h})}_{(\nu)}({\bf k},\omega)},\label{full-charge-longitudinal-part-2}\\
g_{\rm inter}({\bf k},\omega)&=&{1\over 2}e^{i\theta_{\bf k}}\sum_{\nu=1,2}(-1)^{\nu}{1\over \omega-\xi^{(\nu)}_{\bf k}-\Sigma^{({\rm h})}_{(\nu)}({\bf k},\omega)}.~~~~~~~~ \label{full-charge-transverse-part-2}
\end{eqnarray}
\end{subequations}
where $\nu=1,2$, and the mean-field (MF) charge-carrier excitation spectrum $\xi^{(\nu)}_{\bf k}=(-1)^{\nu}Zt\chi|\gamma_{\bf k}|-\mu$, the spin correlation function $\chi=\langle S^{+}_{{\rm A}l}S^{-}_{{\rm B}l+\hat{\eta}}\rangle=\langle S^{+}_{{\rm B} l}S^{-}_{{\rm A}l+\hat{\eta}}\rangle$, $Z=3$ is the number of the NN sites on a honeycomb lattice, $\gamma_{\bf k}=(1/Z)\sum_{\hat{\eta}} e^{-i{\bf k}\cdot {\hat{\eta}}}=e^{i\theta_{\bf k}}|\gamma_{\bf k}|$, while the charge-carrier self-energies are obtained as,
\begin{subequations}\label{self-energy-2}
\begin{eqnarray}
\Sigma^{({\rm h})}_{(1)}({\bf k},\omega)=\Sigma^{({\rm h})}_{\rm intra}({\bf k},\omega)+e^{-i\theta_{\bf k}}\Sigma^{({\rm h})}_{\rm inter}({\bf k},\omega),\\
\Sigma^{({\rm h})}_{(2)}({\bf k},\omega)=\Sigma^{({\rm h})}_{\rm intra}({\bf k},\omega)-e^{-i\theta_{\bf k}}\Sigma^{({\rm h})}_{\rm inter}({\bf k},\omega),
\end{eqnarray}
\end{subequations}
with the corresponding intra-sublattice and inter-sublattice parts of the self-energy, $\Sigma^{({\rm h})}_{\rm intra}({\bf k},\omega)$ and $\Sigma^{({\rm h})}_{\rm inter}({\bf k}, \omega)$, repectively, that are evaluated explicitly in the Appendix \ref{app}. In a doped Mott insulator, the strong correlation effect and the related charge-carrier quasiparticle coherence are closely related to the charge-carrier self-energy $\Sigma^{({\rm h})}_{(\nu)}({\bf k},\omega)$, and then the behavior of the charge-carrier quasiparticle is most fully described in terms of the charge-carrier Green's function (\ref{full-charge-Green-function-2}).

Now we turn to calculate the conductivity of the doped honeycomb-lattice Mott insulators in terms of the full charge-carrier Green's functions (\ref{full-charge-Green-function-2}). The conductivity in the system is expressed as \cite{Mahan81},
\begin{eqnarray}\label{conductivity-1}
\sigma(\omega)=-{{\rm Im}\Pi(\omega)\over\omega},
\end{eqnarray}
with the electron current-current correlation function,
\begin{eqnarray}\label{correlation}
\Pi(\tau-\tau')=-\langle T_{\tau}{\bf j}(\tau)\cdot {\bf j}(\tau')\rangle,
\end{eqnarray}
where ${\bf j}$ is the electron current operator. The external magnetic field can be coupled to the electrons, which are now represented by $C_{\upsilon l\uparrow}= h^{\dagger}_{\upsilon l\uparrow}S^{-}_{\upsilon l}$ and $C_{\upsilon l\downarrow}= h^{\dagger}_{\upsilon l\downarrow}S^{+}_{\upsilon l}$ in the fermion-spin representation (\ref{fermion-spin-theory}). For the calculation of the electron current density, we need to obtain the electron polarization operator, which is a summation over all the particles and their positions, and can be evaluated in the fermion-spin representation as ${\bf P}=-e\sum\limits_{\upsilon l\sigma}{\bf R}_{l}C^{\dagger}_{\upsilon l\sigma}C_{\upsilon l\sigma} = e\sum\limits_{\upsilon l}{\bf R}_{l}h^{\dagger}_{\upsilon l} h_{\upsilon l}$. The electron current operator on the other hand is obtained by evaluating the time-derivative of the polarization operator ${\bf j}= {\partial{\bf P}/\partial t}=(i/\hbar)[H,{\bf P}]$ as \cite{Mahan81},
\begin{widetext}
\begin{eqnarray}\label{tcurpara1}
{\bf j}&=&{iet\over\hbar}\sum\limits_{l\hat{\eta}}\hat{\eta}(h_{{\rm A}l\uparrow}h^{\dagger}_{{\rm B}l+\hat{\eta}\uparrow}S^{+}_{{\rm A}l}S^{-}_{{\rm B}l+\hat{\eta}} +h_{{\rm A} l\downarrow}h^{\dagger}_{{\rm B}l+\hat{\eta}\downarrow}S^{-}_{{\rm A}l}S^{+}_{{\rm B}l+\hat{\eta}}
- h_{{\rm B}l\uparrow}h^{\dagger}_{{\rm A}l+\hat{\eta}\uparrow}S^{+}_{{\rm B}l}S^{-}_{{\rm A}l+\hat{\eta}} -h_{{\rm B}l\downarrow}h^{\dagger}_{{\rm A}l+\hat{\eta}\downarrow} S^{-}_{{\rm B}l}S^{+}_{{\rm A}l+\hat{\eta}}).
\end{eqnarray}
In the fermion-spin approach, this electron current operator (\ref{tcurpara1}) can be decoupled as,
\begin{eqnarray}\label{tcurpara2}
{\bf j}&=&{ie\chi t\over\hbar}\sum\limits_{l\hat{\eta}\sigma}\hat{\eta}(h^{\dagger}_{{\rm A}l+\hat{\eta}\sigma}h_{{\rm B}l\sigma}-h^{\dagger}_{{\rm B}l+\hat{\eta}\sigma}h_{{\rm A}l \sigma})+{ie\phi t\over\hbar}\sum\limits_{l\hat{\eta}}\hat{\eta}(S^{+}_{{\rm A}l}S^{-}_{{\rm B}l+\hat{\eta}}+S^{-}_{{\rm A}l}S^{+}_{{\rm B}l+\hat{\eta}}
-S^{+}_{{\rm B}l}S^{-}_{{\rm A} l+\hat{\eta}}-S^{-}_{{\rm B}l}S^{+}_{{\rm A}l+\hat{\eta}}),
\end{eqnarray}
where $\phi=\langle h^{\dagger}_{{\rm B}l\sigma}h_{{\rm A}l\sigma}\rangle =\langle h^{\dagger}_{{\rm A}l\sigma}h_{{\rm B}l\sigma}\rangle$ is the charge-carrier's particle-hole parameter. The second term in the right-hand side of Eq. (\ref{tcurpara2}) refers to the contribution to the electron current density from the electron spin, however, it has been shown explicitly \cite{Qin14} that $\sum\limits_{l\hat{\eta}}\hat{\eta}(S^{+}_{{\rm A}l}S^{-}_{{\rm B}l+\hat{\eta}}+S^{-}_{{\rm A}l}S^{+}_{{\rm B}l+\hat{\eta}}-S^{+}_{{\rm B}l}S^{-}_{{\rm A}l+\hat{\eta}}-S^{-}_{{\rm B}l}S^{+}_{{\rm A}l+\hat{\eta}})\equiv 0$, i.e., there is no direct contribution to the electron current density from the electron spin, and then the majority contribution to the electron current-current correlation comes from the charge carriers (then the electron charge). However, the strong interplay between the charge carriers and spins has been considered through the spin's order parameters entering in the charge-carrier part of the contribution to the current-current correlation. In this case, the electron current-current correlation function can be evaluated in terms of the full charge carrier Green's function (\ref{full-charge-Green-function-2}) as,
\begin{eqnarray}\label{correlation-1}
\Pi(ip_{m})&=&2(Ze)^{2}{1\over N}\sum_{\bf k}[\gamma_{\rm intra}({\bf k})\Pi^{(g)}_{\rm intra}({\bf k},ip_{m})
+\gamma_{\rm inter}({\bf k})\Pi^{(g)}_{\rm inter}({\bf k},ip_{m})],
\end{eqnarray}
where the intra-sublattice and inter-sublattice current vertices are given by,
\begin{subequations}\label{current-vertices}
\begin{eqnarray}
\gamma_{\rm intra}({\bf k})&=&{4\over 9}(\chi t)^{2}[\sin^{2}(\sqrt{3}k_{y}/2)-\cos(\sqrt{3}k_{y}/2)\cos(3k_{x}/2)+1],\\
\gamma_{\rm inter}({\bf k})&=&-{2\over 9}(\chi t)^{2}[2\cos(2\sqrt{3}k_{y})-2\cos(\sqrt{3}k_{y})+4\cos(3k_{x})\cos(\sqrt{3}k_{y})+4\cos(3k_{x}/2)\cos(\sqrt{3}k_{y}/2)\cos(\sqrt{3}k_{y})\nonumber\\
&-&6\cos(3k_{x}/2)\cos(\sqrt{3}k_{y}/2)+\cos(3k_{x})-7]/[2\cos(\sqrt{3}k_{y})+4\cos(\sqrt{3}k_{y}/2)\cos(3k_{x}/2)+3],
\end{eqnarray}
\end{subequations}
and the intra-sublattice charge-carrier bubble $\Pi_{\rm intra}({\bf k},\omega)$ is obtained from the corresponding intra-sublattice part of the charge-carrier Green's function (\ref{full-charge-longitudinal-part-2}) as,
\begin{eqnarray}\label{bubbl-1}
\Pi^{(g)}_{\rm intra}({\bf k},ip_{m})={1\over\beta}\sum_{i\omega_{n}}g_{\rm intra}({\bf k},i\omega_{n}+ip_{m})g_{\rm intra}({\bf k},i\omega_{n}),
\end{eqnarray}
and is closely related to the quasiparticle scattering within the same sublattice, while the inter-sublattice charge-carrier bubble $\Pi_{\rm inter} ({\bf k},\omega)$ is obtained from the corresponding inter-sublattice part of the charge-carrier Green's function (\ref{full-charge-transverse-part-2}) as,
\begin{eqnarray}\label{bubbl-2}
\Pi^{(g)}_{\rm inter}({\bf k},ip_{m})=e^{-2i\theta_{\bf k}}{1\over\beta}\sum_{i\omega_{n}}g_{\rm inter}({\bf k},i\omega_{n}+ip_{m})g_{\rm inter}({\bf k},i\omega_{n}),~~~~~~~~~~
\end{eqnarray}
and therefore is directly associated with the quasiparticle scattering between the sublattices $A$ and $B$. Substituting the electron current-current correlation function (\ref{correlation-1}) and the intra-sublattice and inter-sublattice charge-carrier bubbles (\ref{bubbl-1}) and (\ref{bubbl-2}) into Eq. (\ref{conductivity-1}), we thus obtain the conductivity as $\sigma(\omega)=\sigma_{\rm intra}(\omega)+\sigma_{\rm inter}(\omega)$, where the intra-sublattice conductivity $\sigma_{\rm intra}(\omega)$ and inter-sublattice conductivity $\sigma_{\rm inter}(\omega)$ can be evaluated explicitly as,
\begin{subequations}\label{conductivity}
\begin{eqnarray}
\sigma_{\rm intra}(\omega)&=&{(Ze)^{2}\over N}\sum_{\bf k}\int^{\infty}_{-\infty}{{\rm d}\omega'\over 2\pi}\gamma_{\rm intra}({\bf k})A^{(\rm h)}_{\rm intra}({\bf k},\omega'+\omega)
A^{(\rm h)}_{\rm intra}({\bf k},\omega'){n_{\rm F}(\omega'+\omega)-n_{\rm F}(\omega')\over\omega}, ~~~~~~~~\label{intra-conductivity}\\
\sigma_{\rm inter}(\omega)&=&{(Ze)^{2}\over N}\sum_{\bf k}e^{-2i\theta_{\bf k}}\int^{\infty}_{-\infty}{{\rm d}\omega'\over 2\pi}\gamma_{\rm inter}({\bf k})A^{(\rm h)}_{\rm inter} ({\bf k},\omega'+\omega)A^{(\rm h)}_{\rm inter}({\bf k},\omega'){n_{\rm F}(\omega'+\omega)-n_{\rm F}(\omega')\over\omega},~~~~~~~~\label{inter-conductivity}
\end{eqnarray}
\end{subequations}
respectively, where the intra-sublattice and inter-sublattice components of the charge-carrier spectral functions are given by $A^{(\rm h)}_{\rm intra}({\bf k},\omega)=-2{\rm Im} g_{\rm intra} ({\bf k},\omega)$ and $A^{(\rm h) }_{\rm inter}({\bf k},\omega)=-2{\rm Im}g_{\rm inter} ({\bf k},\omega)$, respectively, while $n_{\rm F}(\omega)$ is the fermion distribution function.
\end{widetext}

\section{Quantitative characteristics} \label{transport}

\begin{figure*}[t!]
\centering
\includegraphics[scale=0.6]{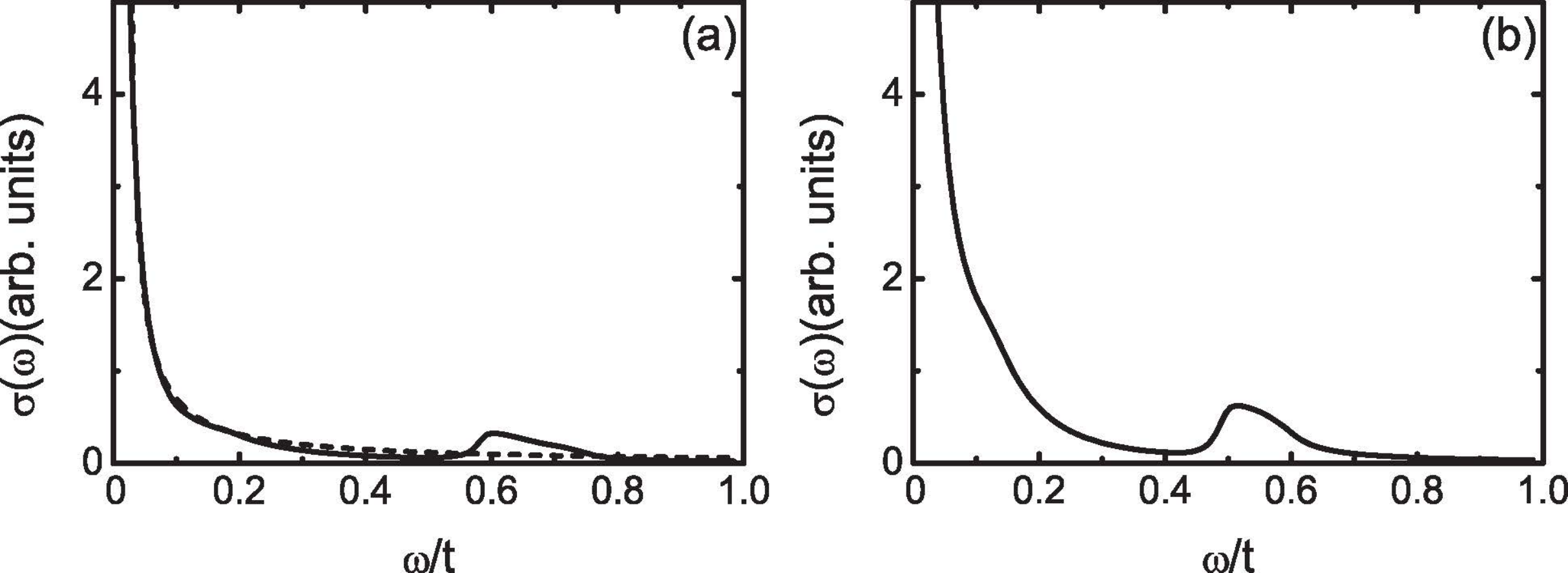}
\caption{The conductivity as a function of energy at (a) $\delta=0.09$ and (b) $\delta=0.15$ with $T=0.002J$ for $t/J=2.5$. The dashed line at (a) is obtained from a numerical fit $\sigma(\omega)=A/(\omega+B)$, with $A\sim 0.058$ and $B\sim -0.017$. \label{conductivity-fit-1}}
\end{figure*}

In Fig. \ref{conductivity-fit-1}, we plot the conductivity $\sigma(\omega)$ as a function of energy at doping levels (a) $\delta=0.09$ and (b) $\delta=0.15$ for parameter $t/J=2.5$ with temperature $T=0.002J$, where the charge $e$ has been set as the unit. It is remarkable that as the case of the doped square-lattice Mott insulators \cite{Qin14}, $\sigma(\omega)$ at the low-doped regime is separated into two bands, with the higher energy band, corresponding to the {\it unusual midinfrared band}, that shows a broad peak around $\omega\sim 0.6t$ for $\delta=0.09$, while the low-energy band forms a sharp peak at $\omega\sim 0$, and deviates strongly from the Drude behavior. To show this deviation from the Drude behavior clearly, we have fitted our present result of $\sigma(\omega)$ for the doping concentration $\delta=0.09$, and the fitted result is also shown in Fig. \ref{conductivity-fit-1}a (dashed line). Obviously, the fitted result of the lower-energy peak decays as $\rightarrow 1/\omega$. This $1/\omega$ decay of the conductivity at low energies is closely related with the linear temperature resistivity, since it reflects an anomalous frequency dependent scattering rate proportional to $\omega$ instead of $\omega^{2}$ as would be expected in the standard Landau Fermi-liquid. Furthermore, as seen from Fig. \ref{conductivity-fit-1}b, although the weight of the midinfrared band increases with the increase of doping, the midinfrared band moves towards to the low-energy non-Drude band. In particular, we find that the low-energy non-Drude peak incorporates with the midinfrared band in the high-doped regime, and then the low-energy Drude type behavior of the conductivity recovers. All these results are qualitatively similar to the case in the doped square-lattice Mott insulators \cite{Qin14}, reflecting that the two-component conductivity is a universal feature for the doped Mott insulators.

\begin{figure}[h!]
\centering
\includegraphics[scale=0.3]{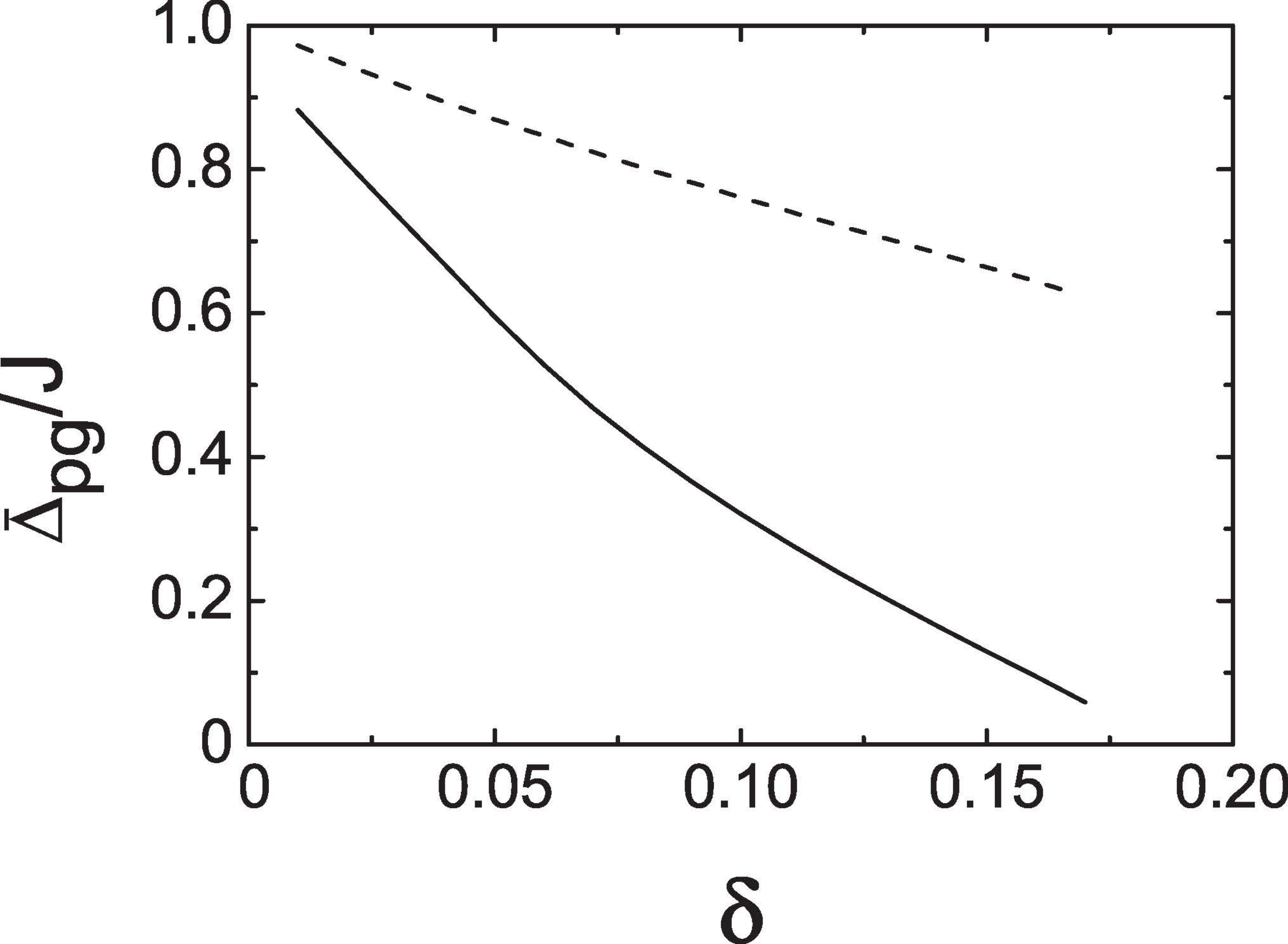}
\caption{The pseudogap parameters $(\bar{\Delta}^{(1)}_{\rm pg})$ (solid line) and $(\bar{\Delta}^{(2)}_{\rm pg})$ (dashed line) as a function of doping with $T=0.002J$ for $t/J=2.5$. \label{pseudogap}}
\end{figure}

In Eq. (\ref{conductivity}), there are two parts of the quasiparticle contribution to the redistribution of the spectral weight in the conductivity $\sigma(\omega)$: the contribution from $\sigma_{\rm intra}(\omega)$ comes from the spectral function obtained in terms of the intra-sublattice part of the charge-carrier Green's function (\ref{full-charge-longitudinal-part-2}), and therefore is closely associated with the quasiparticle scattering within the same sublattice, while the additional contribution from $\sigma_{\rm inter}(\omega)$ originates from the spectral function obtained in terms of the inter-sublattice part of the charge-carrier Green's function (\ref{full-charge-transverse-part-2}), and is closely related to the quasiparticle scattering between the sublattices $A$ and $B$. Within the framework of the fermion-spin theory, the essential physics of the charge transport in the doped honeycomb-lattice Mott insulators is the same as that in the doped square-lattice case \cite{Qin14}, and can be attributed to the emergence of the pseudogap. This follows a fact that there is an intrinsic link between the self-energy and pseudogap. The charge-carrier self-energy $\Sigma^{({\rm h})}_{(\nu)} ({\bf k},\omega)$ in Eq. (\ref{self-energy-2}) also can be rewritten as \cite{Feng15,Feng12},
\begin{eqnarray}\label{self-energy-5}
\Sigma^{({\rm h})}_{(\nu)}({\bf k},\omega)&\approx&{[\bar{\Delta}^{(\nu)}_{\rm pg}({\bf k})]^{2}\over \omega+\varepsilon^{(\nu)}_{\bf k}},
\end{eqnarray}
where $\varepsilon^{(\nu)}_{\bf k}$ is the energy spectrum of $\Sigma^{({\rm h})}_{(\nu)}({\bf k},\omega)$, while $\bar{\Delta}^{(\nu)}_{\rm pg}({\bf k})$ is the pseudogap, and is therefore identified as being a region of the self-energy effect in which the pseudogap suppresses the spectral weight. This pseudogap $\bar{\Delta}^{(\nu)}_{\rm pg}({\bf k})$ and the corresponding energy spectrum $\varepsilon^{(\nu)}_{\bf k}$ can be obtained explicitly in terms of the self-energy $\Sigma^{({\rm h})}_{(\nu)}({\bf k},\omega)$ in Eq. (\ref{self-energy-2}) as $\bar{\Delta}^{(\nu)}_{\rm pg}({\bf k})={L^{(\nu)}_{2}({\bf k})/ \sqrt{L^{(\nu)}_{1}({\bf k})}}$ and $\varepsilon^{(\nu)}_{\bf k}={L^{(\nu)}_{2}({\bf k})/ L^{(\nu)}_{1}({\bf k})}$, respectively, with the functions $L^{(\nu)}_{1}({\bf k})=- \Sigma^{({\rm h})}_{(\nu){\rm o}}({\bf k},\omega=0)$ and $L^{(\nu)}_{2}({\bf k})=\Sigma^{({\rm h} )}_{(\nu){\rm e}}({\bf k},\omega=0)$. In this case, the pseudogap parameter is obtained as $\bar{\Delta}^{(\nu)}_{\rm pg}=(1/N)\sum_{\bf k}\bar{\Delta}^{(\nu)}_{\rm pg}({\bf k})$. In Fig. \ref{pseudogap}, we plot the pseudogap parameters $\bar{\Delta}^{(1)}_{\rm pg}$ (solid line) and $\bar{\Delta}^{(2)}_{\rm pg}$ (dashed line) as a function of doping for $t/J=2.5$ with $T=0.002J$. It is shown clearly that although both $\bar{\Delta}^{(1)}_{\rm pg}$ and $\bar{\Delta}^{(2)}_{\rm pg}$ decrease monotonously with the increase of doping, the decrease of the magnitude with increasing doping is faster in $\bar{\Delta}^{(1)}_{\rm pg}$ than in $\bar{\Delta}^{(2)}_{\rm pg}$. In particular, $\bar{\Delta}^{(1)}_{\rm pg}\approx\bar{\Delta}^{(2)}_{\rm pg}$ near the half-filling. Since both the pseudogaps $\bar{\Delta}^{(1)}_{\rm pg}$ and $\bar{\Delta}^{(2)}_{\rm pg}$ appear in the intra-sublattice and inter-sublattice Green's functions in Eq. (\ref{full-charge-Green-function-2}), they together induce the transfer of the part of the low-energy spectral weight in the conductivity in the low-doped regime to the higher energy region to form the midinfrared band. Moreover, the onset of the region to which the spectral weight is transferred is close to the sum of the pseudogaps $\bar{\Delta}_{\rm pg}\sim \bar{\Delta}^{(1)}_{\rm pg}+\bar{\Delta}^{(2)}_{\rm pg}$, and then the decrease of the pseudogaps with increasing doping therefore leads to a shift of the position of the midinfrared band towards to the low-energy non-Drude peak.

\section{Conclusions}\label{conclusions}

Within the framework of the fermion-spin theory, we have studied the charge transport of the doped Mott insulators on a honeycomb lattice by taking into account the pseudogap effect. Our result shows that the conductivity spectrum in the low-doped regime is separated into a non-Drude peak centered around energy $\omega\sim 0$ followed by a broad midinfrared band. However, the decrease of the pseudogaps with increasing doping leads to a shift of the position of the midinfrared band towards to the low-energy non-Drude peak, and then the low-energy Drude type behavior of the conductivity recovers in the high-doped regime. Incorporating the present result with that obtained in the doped square-lattice Mott insulators \cite{Qin14}, it is thus shown that the two-component conductivity induced by the pseudogap is a universal feature in the doped Mott insulators.

\acknowledgments

The authors would like to thank L\"ulin Kuang and Huaisong Zhao for helpful discussions. This work was supported by the funds from the Ministry of Science and Technology of China under Grant No. 2012CB821403, and the National Natural Science Foundation of China under Grant Nos. 11274044 and 11574032, and YL was supported by the Science Foundation of Hengyang Normal University under Grant No. 13B44, and Hunan Provincial Natural Science Foundation of China under Grant No. 2015JJ3027.

\appendix

\section{Charge-carrier and spin Green's functions in the $t$-$J$ model on a honeycomb lattice}\label{app}

We first define the charge-carrier and spin Green's functions as,
\begin{subequations}\label{Green-functions}
\begin{eqnarray}
g({\bf k},\omega)&=&\left(
\begin{array}{cc}
g_{\rm intra}({\bf k},\omega) & g_{\rm inter}({\bf k},\omega) \\
g^{*}_{\rm inter}({\bf k},\omega) & g_{\rm intra}({\bf k},\omega)
\end{array} \right) \,,\label{charge-Green-function}\\
D({\bf k},\omega)&=&\left(
\begin{array}{cc}
D_{\rm intra}({\bf k},\omega) & D_{\rm inter}({\bf k},\omega) \\
D^{*}_{\rm inter}({\bf k},\omega) & D_{\rm intra}({\bf k},\omega)
\end{array} \right) \,,\label{spin-Green-function-1}\\
D_{\rm (z)}({\bf k},\omega)&=&\left(
\begin{array}{cc}
D_{{\rm (z)}{\rm intra}}({\bf k},\omega) & D_{{\rm (z)}{\rm inter}}({\bf k},\omega) \\
D^{*}_{{\rm (z)}{\rm inter}}({\bf k},\omega) & D_{{\rm (z)}{\rm intra}}({\bf k},\omega)
\end{array} \right) \,.~~~~~~~~~\label{spin-Green-function-2}
\end{eqnarray}
\end{subequations}
Since the spin system in the $t$-$J$ model (\ref{cssham}) is anisotropic away from half-filling, therefore two spin Green's functions $D({\bf k},\omega)$ and $D_{\rm z}({\bf k}, \omega)$ have been defined to describe properly spin propagations \cite{Feng15}. In the MF approximation, the $t$-$J$ model (\ref{cssham}) can be decoupled explicitly as $H_{\rm MFA} =H_{t}+H_{J}-2NZt\phi\chi$, with
\begin{widetext}
\begin{subequations}\label{MFHAM}
\begin{eqnarray}
H_{t}&=&{\chi}t\sum_{l\hat{\eta}\sigma}(h^{\dagger}_{{\rm B}l+\hat{\eta}\sigma}h_{{\rm A}l\sigma}+h^{\dagger}_{{\rm A}l+\hat{\eta}\sigma}h_{{\rm B}l\sigma})
-\mu\sum_{\upsilon l\sigma}h^{\dagger}_{\upsilon l\sigma}h_{\upsilon l\sigma},~~~~~~~\\
H_{J}&=&{1\over 2}J_{{\rm eff}}\sum_{l\hat{\eta}}[\epsilon(S^{+}_{{\rm A}l}S^{-}_{{\rm B}l+\hat{\eta}}+S^{-}_{{\rm A}l}S^{+}_{{\rm B}l+\hat{\eta}}+S^{+}_{{\rm B}l}S^{-}_{{\rm A} l+\hat{\eta}}+S^{-}_{{\rm B}l}S^{+}_{{\rm A}l+\hat{\eta}})+2(S^{\rm z}_{{\rm A}l}S^{\rm z}_{{\rm B}l+\hat{\eta}}+S^{\rm z}_{{\rm B}l}S^{\rm z}_{{\rm A}l+\hat{\eta}})],~~~~~
\end{eqnarray}
\end{subequations}
where $\epsilon=1+2t\phi/J_{{\rm eff}}$. Based on the equation of motion method, it is easy to find the intra-sublattice and inter-sublattice parts of the MF charge-carrier Green's function,
\begin{subequations}\label{MF-charge-Green-function}
\begin{eqnarray}
g^{(0)}_{\rm intra}({\bf k},\omega)&=&{1\over 2}\sum_{\nu=1,2}{1\over \omega-\xi^{(\nu)}_{\bf k}}, \label{MF-charge-longitudinal-part}\\
g^{(0)}_{\rm inter}({\bf k},\omega)&=&{1\over 2}e^{i\theta_{\bf k}}\sum_{\nu=1,2}(-1)^{\nu}{1\over \omega-\xi^{(\nu)}_{\bf k}},\label{MF-charge-transverse-part}
\end{eqnarray}
\end{subequations}
and the intra-sublattice and inter-sublattice parts of the MF spin Green's functions,
\begin{subequations}\label{MF-spin-Green-function}
\begin{eqnarray}
D^{(0)}_{\rm intra}({\bf k},\omega)&=&\sum_{\nu=1,2}{B^{(\nu)}_{\bf k}\over 2\omega^{(\nu)}_{\bf k}}\left ({1\over \omega-\omega^{(\nu)}_{\bf k}}-{1\over\omega+\omega^{(\nu)}_{\bf k}} \right ), \label{MF-spin-longitudinal-part-1}\\
D^{(0)}_{\rm inter}({\bf k},\omega)&=&e^{i\theta_{\bf k}}\sum_{\nu=1,2}(-1)^{\nu}{B^{(\nu)}_{\bf k}\over 2\omega^{(\nu)}_{\bf k}}\left ( {1\over \omega-\omega^{(\nu)}_{\bf k}} - {1\over \omega+\omega^{(\nu)}_{\bf k}}\right ), \label{MF-spin-transverse-part-1} \\
D^{(0)}_{\rm (z)intra}({\bf k},\omega)&=&\sum_{\nu=1,2}{B^{(\nu)}_{{\rm z}{\bf k}}\over 2\omega^{(\nu)}_{{\rm z}{\bf k}}}\left ({1\over \omega-\omega^{(\nu)}_{{\rm z}{\bf k}}}- {1\over \omega+\omega^{(\nu)}_{{\rm z}{\bf k}}}\right ),\label{MF-spin-longitudinal-part-2}\\
D^{(0)}_{\rm (z)inter}({\bf k},\omega)&=&e^{i\theta_{\bf k}}\sum_{\nu=1,2}{B^{(\nu)}_{{\rm z}{\bf k}}\over 2\omega^{(\nu)}_{{\rm z}{\bf k}}}\left ({1\over \omega-\omega^{(\nu)}_{{\rm z} {\bf k}}}- {1\over \omega+\omega^{(\nu)}_{{\rm z}{\bf k}}} \right ), \label{MF-spin-transverse-part-2}
\end{eqnarray}
\end{subequations}
with $B^{(\nu)}_{\bf k}=\lambda[(-1)^{\nu}A_{1}|\gamma_{\bf k}|-A_{2}]$, $B^{(\nu)}_{{\rm z}{\bf k}}=-\lambda\epsilon\chi[1-(-1)^{\nu}|\gamma_{\bf k}|]/2$, $\lambda=2ZJ_{\rm eff}$, $A_{1}=\epsilon\chi^{\rm z}+\chi/2$, $A_{2}=\chi^{\rm z}+\epsilon\chi/2$, the spin correlation function $\chi^{\rm z}=\langle S^{\rm z}_{{\rm A}l} S^{\rm z}_{{\rm B}l+\hat{\eta}} \rangle$, and the MF spin excitation spectra,
\begin{subequations}
\begin{eqnarray}
(\omega^{(\nu)}_{\bf k})^{2}&=&\lambda^{2}[\epsilon\alpha A_{1}(|\gamma|^{2}_{\bf k}-{1\over Z})+{1\over 2}\epsilon^{2}A_{3}+A_{4}] -(-1)^{\nu}\epsilon\lambda^{2}[\alpha A_{2} (1-{1\over Z})+{A_{3}\over 2}+A_{4}]|\gamma_{\bf k}|,~~~~~~~\\
(\omega^{(\nu)}_{{\rm z}{\bf k}})^{2}&=&\epsilon\lambda^{2}[\alpha\chi(|\gamma|^{2}_{\bf k}-{1\over Z})+\epsilon A_{3}]-(-1)^{\nu}\epsilon\lambda^{2}[\alpha\chi(1-{1\over Z} ) +\epsilon A_{3}]|\gamma_{\bf k}|,
\end{eqnarray}
\end{subequations}
where $A_{3}=\alpha C+(1-\alpha)/(2Z)$, $A_{4}=\alpha C_{z}+(1-\alpha)/(4Z)$, and the spin correlation functions $C=(1/Z^{2})\sum_{{\hat{\eta}}{\hat{\eta}'}}\langle S^{+}_{{\rm A} l+{\hat{\eta}}}S^{-}_{{\rm A}l+\hat{\eta}'}\rangle$, $C^{\rm z}=(1/Z^{2})\sum_{{\hat{\eta}}{\hat{\eta}'}}\langle S^{\rm z}_{{\rm A}l+\hat{\eta}} S^{\rm z}_{{\rm A}l+\hat{\eta}'} \rangle$. In order to satisfy the sum rule of the correlation function $\langle S^{+}_{\upsilon l}S^{-}_{\upsilon l}\rangle=1/2$ in the case without an AF long-range order, the important decoupling parameter $\alpha$ has been introduced in the above calculation \cite{Qin14,Feng15}, which can be regarded as the vertex correction.

As in the case of the doped square-lattice Mott insulators \cite{Qin14}, the self-consistent equation that is satisfied by the full charge-carrier Green's function is obtained as \cite{Mahan81},
\begin{eqnarray}\label{full-charge-Green-function}
g({\bf k},\omega)&=&g^{(0)}({\bf k},\omega)+g^{(0)}({\bf k},\omega)\Sigma^{({\rm h})}({\bf k},\omega)g({\bf k},\omega),
\end{eqnarray}
where the intra-sublattice and inter-sublattice parts of the self-energy $\Sigma^{({\rm h})}({\bf k},\omega)$ is obtained from the spin bubble as,
\begin{subequations}\label{self-energy}
\begin{eqnarray}
\Sigma^{({\rm h})}_{\rm intra}({\bf k},i\omega_{n})&=&{1\over N^{2}}\sum_{\bf pq}{|\Lambda_{{\bf p}+{\bf q}+{\bf k}}}|^{2}{1\over\beta}\sum_{ip_{m}}g_{\rm intra}({\bf k}+ {\bf p}, i\omega_{n}+ip_{m})\Pi_{\rm intra}({\bf p},{\bf q},ip_{m}), ~~~~\label{longitudinal-part-self-energy}\\
\Sigma^{({\rm h})}_{\rm inter}({\bf k},i\omega_{n})&=&{1\over N^{2}}\sum_{\bf pq}{\Lambda^{2}_{{\bf p}+{\bf q}+{\bf k}}}{1\over\beta}\sum_{ip_{m},iq_{n}}g^{*}_{\rm inter} ({\bf k}+ {\bf p}, i\omega_{n}+ip_{m})\Pi^{*}_{\rm inter}({\bf p},{\bf q},ip_{m}), ~~~~~~~\label{transverse-part-self-energy}
\end{eqnarray}
\end{subequations}
with $\Lambda_{\bf k}=Zt\gamma_{\bf k}$, and the corresponding intra-sublatticel and inter-sublattice parts of the spin bubble,
\begin{subequations}\label{spin-bubble}
\begin{eqnarray}
\Pi_{\rm intra}({\bf p},{\bf q},ip_{m})&=&{1\over\beta}\sum_{iq_{m}}D^{(0)}_{\rm intra}({\bf q},iq_{m})D^{(0)}_{\rm intra}({\bf p}+{\bf q},ip_{m}+iq_{m}),\\
\Pi_{\rm inter}({\bf p},{\bf q},ip_{m})&=&{1\over\beta}\sum_{iq_{m}}D^{(0)}_{\rm inter}({\bf q},iq_{m})D^{(0)}_{\rm inter}({\bf p}+{\bf q},ip_{m}+iq_{m}).
\end{eqnarray}
\end{subequations}
\end{widetext}
Since the self-energy $\Sigma^{({\rm h})}({\bf k},\omega)$ is not an even function of $\omega$, we follow common practice to separate $\Sigma^{({\rm h})}({\bf k},\omega)$ into the symmetric and antisymmetric parts, i.e., $\Sigma^{({\rm h})}({\bf k},\omega)=\Sigma^{({\rm h})}_{\rm e}({\bf k},\omega)+\omega\Sigma^{({\rm h})}_{\rm o}({\bf k},\omega)$, and then both $\Sigma^{({\rm h})}_{\rm e}({\bf k},\omega)$ and $\Sigma^{({\rm h})}_{\rm o}({\bf k},\omega)$ are an even function of $\omega$. Moreover, $\Sigma^{({\rm h})}_{\rm o}({\bf k}, \omega)$ is directly associated with the charge-carrier quasiparticle coherent weight \cite{Mahan81} as $Z^{-1}_{\rm hF}({\bf k},\omega)=1-{\rm Re}\Sigma^{({\rm h})}_{\rm o}({\bf k}, \omega)$. In the static limit, although $Z_{\rm hF}({\bf k})$ still is a function of momentum, however, the wave vector ${\bf k}$ in $Z_{\rm hF} ({\bf k})$ can be chosen as ${\bf k}_{0}=[2\pi/3,0]$, and then the corresponding intra-sublattice and inter-sublattice parts of the quasiparticle coherent weight can be obtained as,
\begin{eqnarray}\label{self-consistent-equations-1}
Z^{(\nu)-1}_{\rm hF}=1-{\rm Re}\Sigma^{({\rm h})}_{\rm (\nu)o}({\bf k},\omega=0)\mid_{{\bf k}_{0}}.
\end{eqnarray}
In this case, the intra-sublattice and inter-sublattice parts of the full charge-carrier Green's functions in Eq. (\ref{full-charge-Green-function}) can be evaluated explicitly as,
\begin{subequations}\label{full-charge-Green-function-1}
\begin{eqnarray}
g_{\rm intra}({\bf k},\omega)&=&{1\over 2}\sum_{\nu=1,2}{Z^{(\nu)}_{\rm hF}\over \omega-\bar{\xi}^{(\nu)}_{\bf k}}, \label{full-charge-longitudinal-part-1}\\
g_{\rm inter}({\bf k},\omega)&=&{1\over 2}e^{i\theta_{\bf k}}\sum_{\nu=1,2}(-1)^{\nu}{Z^{(\nu)}_{\rm hF}\over\omega-\bar{\xi}^{(\nu)}_{\bf k}},\label{full-charge-transverse-part-1}
\end{eqnarray}
\end{subequations}
where $\bar{\xi}^{(\nu)}_{\bf k}=Z^{(\nu)}_{\rm hF}\xi^{(\nu)}_{\bf k}$. With the help of the Green's functions in Eq. (\ref{full-charge-Green-function-1}) and the spin Green's functions in Eq. (\ref{MF-spin-Green-function}), the intra-sublattice and inter-sublattice parts of the self-energy in Eq. (\ref{self-energy}) can be evaluated explicitly as,
\begin{widetext}
\begin{subequations}\label{self-energy-1}
\begin{eqnarray}
\Sigma^{({\rm h})}_{\rm intra}({\bf k},\omega)&=&{1\over 8N^{2}}\sum_{{\bf pq}\nu\nu'\nu''}{|\Lambda_{{\bf p}+{\bf q}+{\bf k}}}|^{2}Z^{(\nu)}_{\rm hF}{B^{(\nu')}_{\bf q} B^{(\nu'')}_{{\bf q}+{\bf p}} \over\omega^{(\nu')}_{\bf q}\omega^{(\nu'')}_{{\bf q}+{\bf p}}}\left ({F^{(1)}_{{\bf p}{\bf q}{\bf k}}\over \omega+\omega^{(\nu'')}_{{\bf q}+{\bf p}} -\omega^{(\nu')}_{\bf q}- \bar{\xi}^{(\nu)}_{{\bf p}+{\bf k}}}+{F^{(2)}_{{\bf p}{\bf q}{\bf k}}\over \omega-\omega^{(\nu'')}_{{\bf q}+{\bf p}}-\omega^{(\nu')}_{\bf q}- \bar{\xi}^{(\nu)}_{{\bf p}+{\bf k}}}\right . \nonumber\\
&+&\left . {F^{(3)}_{{\bf p}{\bf q}{\bf k}} \over\omega+\omega^{(\nu'')}_{{\bf q}+{\bf p}}+\omega^{(\nu')}_{\bf q}-\bar{\xi}^{(\nu)}_{{\bf p}+{\bf k}}}
+{F^{(4)}_{{\bf p}{\bf q}{\bf k}}\over \omega-\omega^{(\nu'')}_{{\bf q}+{\bf p}}+\omega^{(\nu')}_{\bf q}-\bar{\xi}^{(\nu)}_{{\bf p}+{\bf k}}}\right ), \label{longitudinal-part-self-energy-1}\\
\Sigma^{({\rm h})}_{\rm inter}({\bf k},\omega)&=&{1\over 8N^{2}}\sum_{{\bf pq}\nu\nu'\nu''}(-1)^{\nu+\nu'+\nu''}e^{-i(\theta_{{\bf q}+{\bf p}}+\theta_{\bf q}+\theta_{{\bf p}+{\bf k}} )}{\Lambda^{2}_{{\bf p}+{\bf q}+{\bf k}}}Z^{(\nu)}_{\rm hF} {B^{(\nu')}_{\bf q} B^{(\nu'')}_{{\bf q}+{\bf p}} \over\omega^{(\nu')}_{\bf q}\omega^{(\nu'')}_{{\bf q}+{\bf p}}} \left ({F^{(1)}_{{\bf p}{\bf q}{\bf k}}\over \omega+ \omega^{(\nu'')}_{{\bf q}+{\bf p}}-\omega^{(\nu')}_{\bf q}- \bar{\xi}^{(\nu)}_{{\bf p}+{\bf k}}}\right . \nonumber\\
&+&\left . {F^{(2)}_{{\bf p}{\bf q}{\bf k}}\over \omega-\omega^{(\nu'')}_{{\bf q}+{\bf p}}-\omega^{(\nu')}_{\bf q}-\bar{\xi}^{(\nu)}_{{\bf p}+{\bf k}}}
+{F^{(3)}_{{\bf p}{\bf q}{\bf k}} \over\omega+\omega^{(\nu'')}_{{\bf q}+{\bf p}}+\omega^{(\nu')}_{\bf q}-\bar{\xi}^{(\nu)}_{{\bf p}+{\bf k}}}
+ {F^{(4)}_{{\bf p}{\bf q}{\bf k}}\over \omega-\omega^{(\nu'')}_{{\bf q}+{\bf p}}+\omega^{(\nu')}_{\bf q}-\bar{\xi}^{(\nu)}_{{\bf p}+{\bf k}}}\right ),
\label{transverse-part-self-energy-1}
\end{eqnarray}
\end{subequations}
where $F^{(1)}_{{\bf p}{\bf q}{\bf k}}=n_{\rm B}(\omega^{(\nu'')}_{{\bf q}+{\bf p}})[1+n_{\rm B}(\omega^{(\nu')}_{\bf q})]+n_{\rm F}(\bar{\xi}^{(\nu)}_{{\bf p}+{\bf k}})[n_{\rm B} (\omega^{(\nu')}_{\bf q})-n_{\rm B}(\omega^{(\nu'')}_{{\bf q}+{\bf p}})]$, $F^{(2)}_{{\bf p}{\bf q}{\bf k}}=[1+n_{\rm B}(\omega^{(\nu')}_{\bf q})][1+n_{\rm B}(\omega^{(\nu'')}_{{\bf q} +{\bf p}})]-n_{\rm F}(\bar{\xi}^{(\nu)}_{{\bf p}+{\bf k}})[1+n_{\rm B}(\omega^{(\nu')}_{\bf q})+n_{\rm B}(\omega^{(\nu'')}_{{\bf q}+{\bf p}})]$,
$F^{(3)}_{{\bf p}{\bf q}{\bf k}}=n_{\rm B}(\omega^{(\nu'')}_{{\bf q}+{\bf p}})n_{\rm B}(\omega^{(\nu')}_{\bf q})+n_{\rm F}(\bar{\xi}^{(\nu)}_{{\bf p}+{\bf k}})[1+n_{\rm B} (\omega^{(\nu')}_{\bf q})+n_{\rm B}(\omega^{(\nu'')}_{{\bf q}+{\bf p}})]$, $F^{(4)}_{{\bf p}{\bf q}{\bf k}}=n_{\rm B}(\omega^{(\nu')}_{\bf q})[1+n_{\rm B}(\omega^{(\nu'')}_{{\bf q} +{\bf p}})]-n_{\rm F}(\bar{\xi}^{(\nu)}_{{\bf p}+{\bf k}})[n_{\rm B}(\omega^{(\nu')}_{\bf q})-n_{\rm B}(\omega^{(\nu'')}_{{\bf q}+{\bf p}})]$, and $n_{\rm B}(\omega)$ is the boson distribution function.
\end{widetext}

In this case, the self-consistent equations that are satisfied by the intra-sublattice and inter-sublattice parts of the quasiparticle coherent weight in Eq. (\ref{self-consistent-equations-1}) must be solved simultaneously with following self-consistent equations,
\begin{subequations}\label{self-consistent-equations-2}
\begin{eqnarray}
\delta &=& {1\over 4N}\sum_{\nu,{\bf k}}Z^{(\nu)}_{\rm hF}\left ( 1-{\rm tanh}[{1\over 2}\beta \bar{\xi}^{(\nu)}_{\bf k}]\right ),\\
\phi &=& {1\over 4N}\sum_{\nu,{\bf k}}(-1)^{\nu}|\gamma_{\bf k}|Z^{(\nu)}_{\rm hF}\left ( 1-{\rm tanh}[{1\over 2}\beta\bar{\xi}^{(\nu)}_{\bf k}]\right ),~~~~~~~~\\
{1\over 2} &=& {1\over 2N}\sum_{\nu,{\bf k}}{B^{(\nu)}_{\bf k}\over\omega^{(\nu)}_{\bf k}}{\rm coth} [{1\over 2}\beta\omega^{(\nu)}_{\bf k}],\\
\chi &=& {1\over 2N}\sum_{\nu,{\bf k}}(-1)^{\nu}|\gamma_{\bf k}|{B^{(\nu)}_{\bf k}\over\omega^{(\nu)}_{\bf k}}{\rm coth} [{1\over 2}\beta\omega^{(\nu)}_{\bf k}],
\end{eqnarray}
\begin{eqnarray}
C &=& {1\over 2N}\sum_{\nu,{\bf k}}|\gamma_{\bf k}|^{2}{B^{(\nu)}_{\bf k}\over\omega^{(\nu)}_{\bf k}}{\rm coth} [{1\over 2}\beta\omega^{(\nu)}_{\bf k}],\\
\chi^{\rm z} &=& {1\over 2N}\sum_{\nu,{\bf k}}(-1)^{\nu}|\gamma_{\bf k}|{B^{(\nu)}_{{\rm z}{\bf k}}\over\omega^{(\nu)}_{{\rm z}{\bf k}}}{\rm coth} [{1\over 2}\beta \omega^{(\nu)}_{{\rm z}{\bf k}}],~~~~~~\\
C^{\rm z}&=&{1\over 2N}\sum_{\nu,{\bf k}}|\gamma_{\bf k}|^{2}{B^{(\nu)}_{{\rm z}{\bf k}}\over\omega^{(\nu)}_{{\rm z}{\bf k}}}{\rm coth} [{1\over 2}\beta\omega^{(\nu)}_{{\rm z}{\bf k} }],
\end{eqnarray}
\end{subequations}
then all the order parameters, the decoupling parameter $\alpha$, and the chemical potential $\mu$ are determined by the self-consistent calculation without using any adjustable parameters. With the above self-consistent calculation of the order parameters, the intra-sublattice and inter-sublattice parts of the full charge-carrier Green's functions can be evaluated explicitly as,
\begin{subequations}\label{full-charge-Green-function-5}
\begin{eqnarray}
g_{\rm intra}({\bf k},\omega)&=&{1\over 2}\sum_{\nu}{1\over \omega-\xi^{(\nu)}_{\bf k}-\Sigma^{({\rm h})}_{(\nu)}({\bf k},\omega)},\\
g_{\rm inter}({\bf k},\omega)&=&{1\over 2}e^{i\theta_{\bf k}}\sum_{\nu}(-1)^{\nu}{1\over \omega-\xi^{(\nu)}_{\bf k}-\Sigma^{({\rm h})}_{(\nu)}({\bf k},\omega)},~~~~~~~~~~
\end{eqnarray}
\end{subequations}
as quoted in Eq. (\ref{full-charge-Green-function-2}).

\end{document}